\documentstyle[preprint,aps,epsfig,psfig]{revtex}
\begin{document}


\title{\Large\bf 
Modification of Kawai Model about the Mixing of the Pseudoscalar Mesons
}
\vspace{1cm}

\author{ \small De-Min Li$^1$\footnote{E-mail: lidm@hptc5.ihep.ac.cn}, 
~~Hong Yu$^{1,2,3}$,~ Qi-Xing Shen$^{1,2,3}$\\ 
\small$^1$Institute of High Energy Physics, Chinese Academy of Sciences,\\ 
\small P.O.Box $918~(4)$, Beijing $100039$, China\footnote{Mailing
address}\\
\small $^2$CCAST (World Lab), P.O.Box 8730, Beijing $100080$, China\\
\small$^3$Institute of Theoretical Physics, Chinese Academy of Sciences, 
Beijing 100080, China } 
\maketitle
\vspace*{0.3cm}

\begin{abstract}
The Kawai model describing the glueball-quarkonia mixing is modified. The
mixing of $\eta$, $\eta^\prime$ and $\eta(1410)$ is re-investigated based 
on the modified Kawai model. The glueball-quarkonia content of the three
states is determined from a fit to the data of the electromagnetic decays 
involving $\eta$, $\eta^\prime$. Some predictions about the 
electromagnetic decays involving $\eta(1410)$ are presented. 
\end{abstract}

\vspace{1.5cm}



\newpage

\section*{I. Introduction}
\indent

The $0^{-}$ ground state nonet is one of the best established $q\bar{q}$
multiplets. The  
isotriplet $\pi(1300)$ and the isodoublet $K(1460)$ of the $0^{-}$ first
radial excitation have been established\cite{REP} and the $\eta(1295)$
can be identified as the first radial excitation of $\eta$\cite{RMP,BAR}.
In addition, $\eta(1440)$ has been resolved into two states: $\eta(1490)$
and $\eta(1410)$\cite{AUS,BAI,BER}(let $\eta^{\prime\prime}$ stand for 
the $\eta(1410)$ below). The former has been interpreted as the
mainly $s \bar{s}$ radial excitation of $\eta^\prime$\cite{RMP,BAR,CLO,KIT}
and the latter seems a spurious state, which is argued to be a mainly
glueball, possibly mixed with $q\bar{q}$ states\cite{CLO,KIT}.

In general, states with the same isospin-spin-parity $IJ^{PC}$ and
additive quantum numbers can mix. 
The fact 
that $M_{\eta(1295)}\approx M_{\pi(1300)}$\cite{PDG} implies that 
$\eta(1490)$ and $\eta(1295)$ are almost ideal mixing. 
Therefore, the possibility of mixing of ground states and radial 
excitations can be ignored, then one can focus on the mixing of $\eta$, 
$\eta^\prime$ and $\eta^{\prime\prime}$. 
The mixing of $\eta$, $\eta^\prime$ and $\eta^{\prime\prime}$ has been
discussed in Ref.\cite{KAW} based on the mass-squared matrix
\begin{equation}
M^2=\left(\begin{array}{ccc}
M^2_N+rA_1&\sqrt{r}A_1&\sqrt{r}A_2\\
\sqrt{r}A_1&M^2_{S}+A_1&A_2\\
\sqrt{r}A_2&A_2&M^2_{G_0}+A_3
 \end{array}\right)
\end{equation}
with the $|N\rangle=|u\bar{u}+d\bar{d}\rangle/\sqrt{2}$,
$|S\rangle=|s\bar{s}\rangle$
and $|G_0\rangle=|gg\rangle$
basis\footnote{Here, $A_1$, $A_2$, $A_3$ and $r$ respetively correspond 
to $\lambda^2_S$, $\lambda_S\lambda_G$, $\lambda^2_G$ and
$2\lambda^2_N/\lambda^2_S$ employed in Ref.\cite{KAW}}, where $M_N$,
$M_{S}$ 
and $M_{G_0}$ are the masses of primitive (unmixed)
$|N\rangle$, $|S\rangle$ and $|G_0\rangle$, respectively; $A_1$, $A_2$,
$A_3=A^2_2/A_1$ describe the transitions between strangeonium and 
strangeonium, between strangeonium and gluonium, and between gluonium and
gluonium, respectively. $r$ describes
the effect of flavor-dependent transition taking
into account the possibility that the nonstrange quarkonia and
strange quarkonia system have the different wave functions at the origin
as the result of the different mass (in $SU(3)$ limit, $r=2$). The
eigenvalues of $M^2$ are
$M^2_{\eta}$, $M^2_{\eta^\prime}$ and $M^2_{\eta^{\prime\prime}}$, the
masses square of the physical states $\eta$, $\eta^\prime$ and
$\eta^{\prime\prime}$, respectively.

However, we believe this mixing model should be modified for the
pseudoscalar
mesons. In Ref\cite{KAW}, it is pointed out that
$M^2_{G_0}\simeq 2\langle k^2_T\rangle$ for a digluon-ball, where
$\langle k^2_T\rangle$ is the transverse momentum fluctuation of
the constituent gluons, and that $A_3$ is considered as the additional
contribution to the matrix $M^2$ due to the transition between gluonium
and gluonium. In the viewpoint of lattice QCD, the value of 
$M_{G_0}$ would be related to the prediction about the
mass of the pseudoscalar
glueball in quenched approximation since $A_3$ at least can contain the 
contribution arising from the transitions of $|G_0\rangle$ to a quark pair
and back to $|G_0\rangle$.
However, based on Eq. (1) $M_{G_0}$ is
determined to be the value of
about $1.3$ GeV\cite{KAW}, which is obviously inconsistent with
$2.56\pm 0.13$ GeV\cite{MOR}, the mass of pseudoscalar glueball
predicted by lattice QCD in quenched approximation. Furthermore, in
the presence of $A_3=A^2_2/A_1$, if one
restricts $M_{G_0}$ to be comparable with the prediction given by lattice
QCD in quenched approximation for the pseudoscalar glueball mass in the 
matrix $M^2$ (i.e., $M_{G_0}>2$ GeV) and assumes the eigenvalues of $M^2$
are the masses square of $\eta$, $\eta^\prime$ and
$\eta^{\prime\prime}$, respectively, based on Eqs. (6)$\sim$(8), one can
have $A^2_2<0$ which would cause the matrix $M^2$ to be a non-hermitian 
matrix. In fact, in the pseudoscalar mesons sector,
$A_1$, $A_2$ and $A_3$ should be nonperturbative effect, and the relation
of $A_1$,
$A_2$ and $A_3$ is completely unknown in principle, therefore there is no
convincing reason to expect that the relation of $A_1$, $A_2$ and
$A_3$ should behave as $A_3=A^2_2/A_1$. In this work, we shall relate
$M_{G_0}$ to the prediction of the pseudoscalar glueball mass given by
lattice QCD in quenched approximation
and consider $A_3$ as a free parameter describing the sum of all
fermion-loop corrections to the quenched prediction of the pseudoscalar
glueball mass. 

\section*{II. Mixing of $\eta$, $\eta^\prime$ and $\eta^{\prime\prime}$ 
based on the modified  Kawai Model}

If $A_3$ is considered as a free 
parameter rather than $A_3=A^2_2/A_1$ as usual in Ref.\cite{KAW},
diagonolizing the matrix $M^2$, one can get
\begin{equation} UM^2U^\dagger=\left(\begin{array}{ccc}
 M^2_{\eta{\prime\prime}}&0&0\\
0&M^2_{\eta^\prime}&0\\ 0&0&M^2_{\eta} 
\end{array}\right), 
\end{equation}
where
\begin{eqnarray} 
U=\left(\begin{array}{ccc}
x_{\eta^{\prime\prime}}&y_{\eta^{\prime\prime}}&z_{\eta^{\prime\prime}}\\
x_{\eta^\prime}&y_{\eta^\prime}&z_{\eta^\prime}\\
x_\eta&y_\eta&z_\eta
\end{array}\right)
\end{eqnarray}
and
\begin{eqnarray}
&&x_i=\sqrt{r}(M^2_i-M^2_S)(A^2_2-A_1A_3+A_1M^2_i-A_1M^2_{G_0})/f_i,
\nonumber\\
&&y_i=(M^2_i-M^2_N)(A^2_2-A_1A_3+A_1M^2_i-A_1M^2_{G_0})/f_i,
\\
&&z_i=(M^2_i-M^2_N)(M^2_i-M^2_S)A_2/f_i,
\nonumber
\end{eqnarray} 
with 
\begin{eqnarray}
f_i=
&&\{r[(M^2_i-M^2_S)(A^2_2-A_1A_3+A_1M^2_i-A_1M^2_{G_0})]^2
\nonumber\\
&&+[(M^2_i-M^2_N)(A^2_2-A_1A_3+A_1M^2_i-A_1M^2_{G_0})]^2
\nonumber\\
&&+[(M^2_i-M^2_N)(M^2_i-M^2_S)A_2]^2\}^{\frac{1}{2}},
\nonumber
\end{eqnarray}
$i$=$\eta^{\prime\prime}$, $\eta^\prime$ and $\eta$. 
The physical states $|\eta\rangle$, $|\eta^\prime\rangle$ and 
$|\eta^{\prime\prime}\rangle$
can be read as 
\begin{equation} 
\left(\begin{array}{c} |\eta^{\prime\prime}\rangle\\
|\eta^\prime\rangle\\ 
|\eta\rangle
 \end{array}\right) 
=U\left(\begin{array}{c} |N\rangle\\ |S\rangle\\ |G_0\rangle
\end{array}\right).  
\end{equation} 
From Eq. (2), one can have 
\begin{eqnarray}
&&M^2_{\eta^{\prime\prime}}M^2_{\eta^{\prime}}M^2_{\eta}=(A_3 +
M^2_{G_0})(A_1M^2_N +M^2_NM^2_S + A_1M^2_Sr)-A^2_2(M^2_N +M^2_Sr),\\
&& M^2_{\eta^{\prime\prime}}M^2_{\eta^{\prime}} +
M^2_{\eta^{\prime\prime}}M^2_{\eta} 
+ M^2_{\eta^{\prime}}M^2_{\eta}=A_3M^2_N +M^2_{G_0}M^2_N
+A_3M^2_S+M^2_{G_0}M^2_S+M^2_NM^2_S-A^2_2(1+r)\nonumber\\
&&~~~~~~~~~~~~~~~~~~~~~~~~~~~~~~~~~~~~~~~~+A_1(A_3 + M^2_{G_0} +
M^2_N
+ A_3r +M^2_{G_0}r +M^2_Sr),\\
&&M^2_{\eta^{\prime\prime}}+M^2_{\eta^{\prime}}+
M^2_{\eta}=M^2_N+M^2_S+M^2_{G_0}+rA_1+A_1+A3.
\end{eqnarray}

For the electromagnetic decays involving
$\eta$, $\eta^\prime$ and $\eta^{\prime\prime}$, based on Eq. 
(5), performing an elementary $SU(3)$ calculation\cite{SEI,HAB,FCL}, 
one can obtain the following equations:
\begin{eqnarray}
&&\frac{\Gamma(\eta\rightarrow\gamma\gamma)}
{\Gamma(\pi^0\rightarrow\gamma\gamma)}=\frac{1}{9}
\left(\frac{M_\eta}{M_{\pi^0}}\right)^3
(5x_\eta+\sqrt{2}y_\eta)^2,\\
&&\frac{\Gamma(\eta^\prime\rightarrow\gamma\gamma)}
{\Gamma(\pi^0\rightarrow\gamma\gamma)}=\frac{1}{9}
\left(\frac{M_{\eta^\prime}}{M_{\pi^0}}\right)^3
(5x_{\eta^\prime}+\sqrt{2}y_{\eta^\prime})^2,\\
&&\frac{\Gamma(\rho\rightarrow\eta\gamma)}
{\Gamma(\omega\rightarrow\pi^0\gamma)}=
\left[
\frac{(M^2_\rho-M^2_\eta)M_\omega}
{(M^2_\omega-M^2_{\pi^0})M_\rho}
\right]^3x^2_\eta,\\
&&\frac{\Gamma(\eta^\prime\rightarrow\rho\gamma)} 
{\Gamma(\omega\rightarrow\pi^0\gamma)}=
3\left[
\frac{(M^2_{\eta^\prime}-M^2_\rho)M_\omega}
{(M^2_\omega-M^2_{\pi^0})M_{\eta^\prime}}
\right]^3x^2_{\eta^\prime},\\
&&\frac{\Gamma(\phi\rightarrow\eta\gamma)}
{\Gamma(\omega\rightarrow\pi^0\gamma)}
=\frac{4}{9}\frac{m^2_u}{m^2_s}\left[
\frac{(M^2_\phi-M^2_\eta)M_\omega}{(M^2_\omega-M^2_{\pi^0})M_\phi}
\right]^3 y^2_\eta,\\
&&\frac{\Gamma(\phi\rightarrow\eta^\prime\gamma)}
{\Gamma(\omega\rightarrow\pi^0\gamma)}=
\frac{4}{9}\frac{m^2_u}{m^2_s}
\left[
\frac{(M^2_\phi-M^2_{\eta^\prime})M_\omega}{(M^2_\omega-M^2_{\pi^0})M_\phi}
\right]^3y^2_{\eta^\prime},\\
&&\frac{\Gamma(J/\psi\rightarrow\rho\eta)}
{\Gamma(J/\psi\rightarrow\omega\pi^0)}
=\left[\frac{\sqrt{[M^2_{J/\psi}-(M_\rho+M_\eta)^2][
M^2_{J/\psi}-(M_\rho-M_\eta)^2]}}
{\sqrt{[M^2_{J/\psi}-(M_\omega+M_{\pi^0})^2][
M^2_{J/\psi}-(M_\omega-M_{\pi^0})^2]}}\right]^3x^2_\eta,\\
&&\frac{\Gamma(J/\psi\rightarrow\rho\eta^\prime)}
{\Gamma(J/\psi\rightarrow\omega\pi^0)}
=\left[\frac{\sqrt{[M^2_{J/\psi}-(M_\rho+M_{\eta^\prime})^2][
M^2_{J/\psi}-(M_\rho-M_{\eta^\prime})^2]}}   
{\sqrt{[M^2_{J/\psi}-(M_\omega+M_{\pi^0})^2][
M^2_{J/\psi}-(M_\omega-M_{\pi^0})^2]}}\right]^3x^2_{\eta^\prime},\\
&&\frac{\Gamma(\eta^{\prime\prime}\rightarrow\gamma\gamma)}
{\Gamma(\pi^0\rightarrow\gamma\gamma)}=\frac{1}{9}
\left(\frac{M_{\eta^{\prime\prime}}}{M_{\pi^0}}\right)^3
(5x_{\eta^{\prime\prime}}+\sqrt{2}y_{\eta^{\prime\prime}})^2,\\
&&\frac{\Gamma(\eta^{\prime\prime}\rightarrow\rho\gamma)}
{\Gamma(\omega\rightarrow\pi^0\gamma)}=
3\left[
\frac{(M^2_{\eta^{\prime\prime}}-M^2_\rho)M_\omega}
{(M^2_\omega-M^2_{\pi^0})M_{\eta^{\prime\prime}}}  
\right]^3x^2_{\eta^{\prime\prime}},\\
&&\frac{\Gamma(\eta^{\prime\prime}\rightarrow\omega\gamma)}
{\Gamma(\omega\rightarrow\pi^0\gamma)}=
\frac{1}{3}\left[
\frac{(M^2_{\eta^{\prime\prime}}-M^2_\omega)M_\omega}
{(M^2_\omega-M^2_{\pi^0})M_{\eta^{\prime\prime}}}
\right]^3x^2_{\eta^{\prime\prime}},\\
&&\frac{\Gamma(\eta^{\prime\prime}\rightarrow\phi\gamma)}
{\Gamma(\omega\rightarrow\pi^0\gamma)}=
\frac{4}{9}\frac{m^2_u}{m^2_s}\left[
\frac{(M^2_{\eta^{\prime\prime}}-M^2_\phi)M_\omega}
{(M^2_\omega-M^2_{\pi^0})M_{\eta^{\prime\prime}}}
\right]^3y^2_{\eta^{\prime\prime}},\\
&&\frac{\Gamma(J/\psi\rightarrow\rho\eta^{\prime\prime})}
{\Gamma(J/\psi\rightarrow\omega\pi^0)}
=\left[\frac{\sqrt{[M^2_{J/\psi}-(M_\rho+M_{\eta^{\prime\prime}})^2][
M^2_{J/\psi}-(M_\rho-M_{\eta^{\prime\prime}})^2]}}   
{\sqrt{[M^2_{J/\psi}-(M_\omega+M_{\pi^0})^2][
M^2_{J/\psi}-(M_\omega-M_{\pi^0})^2]}}\right]^3
x^2_{\eta^{\prime\prime}},
\end{eqnarray}
where $M_\rho$, $M_\omega$, $M_\phi$ and $M_{J/\psi}$ are the masses of 
$\rho$, $\omega$, $\phi$ and $J/\psi$, respectively; $m_u$ and $m_s$ are
the
masses of the constituent quark $u$ and $d$, respectively.  

\section*{III. Fit results}

In Eq. (4), we take $M_{G_0}=2.56\pm0.13$ GeV\cite{MOR} and assume 
$M_N=M_{\pi^0}$\cite{KAW,FEL}, then $M_S$ can be obtained from 
Gell-Mann-Okubo mass formula\cite{OKU}
\begin{equation}
M^2_{S}=2M^2_K-M^2_N,
\end{equation}
where $M^2_K=(M^2_{K^{\pm}}+M^2_{K^0})/2$, and
$M_{K^{\pm}}$, $M_{\pi^0}$ are the masses of pseudoscalar mesons
$K^{\pm}$ and $\pi^0$, respectively. 
Apart from $M_{G_0}$, $M_N$, $M_S$ and the masses
of the observed mesons used in this paper (All the
values of mass of the observed mesons used in this paper
are taken from Particle Data Group 98\cite{PDG} except for
$M_{\eta^{\prime\prime}}=1416\pm 2$ MeV\cite{BER}), we take 
the experimental data of Eqs. (9)$\sim$(16)\cite{PDG} (see TABLE I) and
$m_u/m_s=0.642$\cite{xxx} as input. In this way, we
use the 11 equations, 
(6)$\sim$(16), to determine the 4 unknown parameters in Eqs. (4), $A_1$,
$A_2$, $A_3$ and $r$. The parameters are determined as
$A_1=0.2493$ GeV$^2$,~$A_2=-0.2386$ GeV$^2$,~$A_3=-4.8105$
GeV$^2$ and $r=2.9605$ with $\chi^2/d.o.f$(the $\chi^2$ per
degree of freedom)$=1.99/7$. Based on the values of above parameters, the 
matrix $M^2$ remains hermitian, and from Eqs. (3) and (4), the unitary 
matrix $U$ can be given by \begin{equation}
U=\left(\begin{array}{ccc}
x_{\eta^{\prime\prime}}&y_{\eta^{\prime\prime}}&z_{\eta^{\prime\prime}}\\
x_{\eta^\prime}&y_{\eta^\prime}&z_{\eta^\prime}\\
x_\eta&y_\eta&z_\eta
\end{array}\right)=
\left(\begin{array}{ccc}
~0.3879&~0.2924&-0.8741\\
-0.5693&-0.6698&-0.4766\\
~0.7249&-0.6825&~0.0933
\end{array}\right).
\end{equation}
From Eq. (5), the physical states $\eta$, $\eta^\prime$ and
$\eta^{\prime\prime}$ can be read as
\begin{eqnarray}
&&|\eta^{\prime\prime}\rangle=0.3879|N\rangle+0.2924|S\rangle-0.8741|G_0\rangle,
\nonumber\\
&&|\eta^\prime\rangle=-0.5693|N\rangle-0.6698|S\rangle-0.4766|G_0\rangle,
\\
&&|\eta\rangle=0.7249|N\rangle-0.6825|S\rangle+0.0933|G_0\rangle.
\nonumber
\end{eqnarray}
The fit results of Eqs. (9)$\sim$(21) are shown in TABLE I.

Eq. (24) shows that $\eta^{\prime\prime}$ ($\eta^\prime$, $\eta$) 
contains
about $15\%$ ($32.4\%$, $52.5\%$) $(u\bar{u}+d\bar{d})/\sqrt{2}$ component,
$8.5\%$
($44.9\%$, $46.6\%$) $s\bar{s}$ component and
$76.5\%$
($22.7\%$, $0.9\%$) glueball component, which supports the argument that
$\eta^{\prime\prime}$ is a mixed $q\bar{q}$ glueball having a large
glueball component\cite{CLO,KIT}. Eq. (24) also shows  
that the interference between $|N\rangle$ and
$|S\rangle$ is
constructive for
$\eta^{\prime\prime}$ and $\eta^\prime$ while destructive for $\eta$
and that the
interference between $|S\rangle$ and $|G_0\rangle$ is destructive for
$\eta^{\prime\prime}$ and $\eta$ while constructive for $\eta^\prime$. 
Furthermore, the value of $A_3$ shows that fermion-loop corrections to
the mass of
the pseudoscalar glueball obtained in quenched approximation is quite
large, which disagrees with that the quenched prediction agrees
with the full QCD (unquenched) value to within $10\%$\cite{AOK}.

\section*{IV. Summary and Conclusions}
We modify Kawai model and re-investigate the mixing of $\eta$, 
$\eta^\prime$ and $\eta^{\prime\prime}$ based on the modified model. The 
glueball-quarkonia content of the three states is determined from a fit 
to the data of the electromagnetic decays involving $\eta$, 
$\eta^\prime$. Some predictions about the electromagnetic decays
involving $\eta(1410)$ are presented. Our 
conclusions are as follows:

1). In the presence of $A_3=A^2_2/A_1$, in order to make the 
matrix $M^2$ remain hermitian, the mass of the pseudoscalar glueball in
quenched approximation, $M_{G_0}$, would be less than 2 GeV, which is 
inconsistent with the prediction given by lattice QCD in quenched 
approximation. However, in the absence of $A_3=A^2_2/A_1$, not only can 
$M_{G_0}$ be related to the prediction given by lattice QCD in 
quenched approximation but also 
the matrix $M^2$ remains hermitian.

2). $\eta$ is dominantly a $q\bar{q}$ meson ($52.5\%$ 
$(u\bar{u}+d\bar{d})/\sqrt{2}$ and $46.6\%$ $s\bar{s}$) with a negligible 
glueball component ($0.9\%$). $\eta^\prime$ is dominantly a $q\bar{q}$
meson ($32.4\%$
$(u\bar{u}+d\bar{d})/\sqrt{2}$ and $44.9\%$ $s\bar{s}$) with a quite 
large admixture of glueball ($22.7\%$). $\eta^{\prime\prime}$ is 
dominantly a glueball ($76.5\%$) with a admixture of 
$(u\bar{u}+d\bar{d})/\sqrt{2}$ ($15\%$) and $s\bar{s}$ ($8.5\%$). 

3). The interference between $|N\rangle$ and $|S\rangle$ is constructive
for
$\eta^{\prime\prime}$ and $\eta^\prime$ while destructive for $\eta$. The
interference between $|S\rangle$ and $|G_0\rangle$ is destructive for
$\eta^{\prime\prime}$ and $\eta$ while constructive for $\eta^\prime$.

4). For the mass of the pseudoscalar glueball, the fermion-loop
corrections to the prediction given by lattice QCD in quenched 
approximation is quite large, which disagrees with the argument that 
the quenched prediction agrees
with the full QCD (unquenched) value to within $10\%$\cite{AOK}.

\section*{V. Acknowledgments}
This project is supported by the National Natural Science Foundation of
China under Grant No. 19991487, No. 19677205, No. 19835060, and Grant No.
LWTZ-1298 of
the Chinese Academy of Sciences.

\begin{table}

\begin{tabular}{cccccc}
Decay Modes&Fit&Exp.\cite{PDG}&Decay
Modes&Fit&Exp.\cite{PDG}\\\hline
$\frac{\Gamma(\eta\rightarrow\gamma\gamma)}
{\Gamma(\pi^0\rightarrow\gamma\gamma)}$&$52.36$&$58.46\pm9.03$&
$\frac{\Gamma(\eta^{\prime}\rightarrow\gamma\gamma)}
{\Gamma(\pi^0\rightarrow\gamma\gamma)}$&571.07&$540.78\pm104.44$\\\hline
$\frac{\Gamma(\eta^{\prime\prime}\rightarrow\gamma\gamma)}
{\Gamma(\pi^0\rightarrow\gamma\gamma)}$&709.90&&
$\frac{\Gamma(\rho\rightarrow\eta\gamma)}
{\Gamma(\omega\rightarrow\pi^0\gamma)}$&$0.067$&$0.051\pm0.023$\\\hline
$\frac{\Gamma(\eta^\prime\rightarrow\rho\gamma)}
{\Gamma(\omega\rightarrow\pi^0\gamma)}$&$0.087$&$0.086\pm0.016$&
$\frac{\Gamma(\eta^{\prime\prime}\rightarrow\rho\gamma)}
{\Gamma(\omega\rightarrow\pi^0\gamma)}$&1.025&\\\hline  
$\frac{\Gamma(\eta^{\prime\prime}\rightarrow\omega\gamma)}
{\Gamma(\omega\rightarrow\pi^0\gamma)}$&0.110&&
$\frac{\Gamma(\eta^{\prime\prime}\rightarrow\phi\gamma)}
{\Gamma(\omega\rightarrow\pi^0\gamma)}$&0.011&\\\hline
$\frac{\Gamma(\phi\rightarrow\eta\gamma)}
{\Gamma(\omega\rightarrow\pi^0\gamma)}$&$0.075$&$0.078\pm0.010$&
$\frac{\Gamma(\phi\rightarrow\eta^\prime\gamma)}
{\Gamma(\omega\rightarrow\pi^0\gamma)}$&0.0003&$0.0007\pm0.0005$\\\hline 
$\frac{\Gamma(J/\psi\rightarrow\rho\eta)}
{\Gamma(J/\psi\rightarrow\omega\pi^0)}$&$0.474$&$0.460\pm0.120$&
$\frac{\Gamma(J/\psi\rightarrow\rho\eta^\prime)}
{\Gamma(J/\psi\rightarrow\omega\pi^0)}$&$0.226$&$0.250\pm0.079$\\\hline
$\frac{\Gamma(J/\psi\rightarrow\rho\eta^{\prime\prime})}
{\Gamma(J/\psi\rightarrow\omega\pi^0)}$&0.061&
\end{tabular}
\vspace{1cm}
\caption{The fit results as well as the experimental data of the
electromagnetic decays involving $\eta$,
$\eta^\prime$ and $\eta^{\prime\prime}$.}
\end{table}

\end{document}